\documentclass[aps,twocolumn,a4paper,superscriptaddress]{revtex4}
\usepackage{tensor}
\usepackage{graphicx}
\usepackage{amsmath}
\usepackage{amssymb}
\usepackage{enumerate}
\usepackage{subfigure}
\usepackage{tabularx}
\usepackage[colorlinks=true, pdfstartview=FitV, linkcolor=blue, citecolor=red, urlcolor=black, breaklinks=true]{hyperref}

\newcommand{\be}{\begin{equation}}
\newcommand{\ee}{\end{equation}}
\newcommand{\ben}{\begin{eqnarray}}
\newcommand{\een}{\end{eqnarray}}
\newcommand{\bes}{\begin{subequations}}
\newcommand{\ees}{\end{subequations}}
\def\bal#1\eal{\begin{align}#1\end{align}}
\newcommand{\vphi}{\varphi}

\newcommand{\LL}{{\mathcal L}}

\begin{document}
\title{First order framework for gauge k-vortices}
\author{D. Bazeia}\affiliation{Departamento de F\'\i sica, Universidade Federal da Para\'\i ba, 58051-970 Jo\~ao Pessoa, PB, Brazil}
\author{L. Losano}\affiliation{Departamento de F\'\i sica, Universidade Federal da Para\'\i ba, 58051-970 Jo\~ao Pessoa, PB, Brazil}
\author{M.A. Marques}\affiliation{Departamento de F\'\i sica, Universidade Federal da Para\'\i ba, 58051-970 Jo\~ao Pessoa, PB, Brazil}
\author{R. Menezes}\affiliation{Departamento de Ci\^encias Exatas, Universidade Federal da Para\'{\i}ba, 58297-000 Rio Tinto, PB, Brazil}\affiliation{Departamento de F\'\i sica, Universidade Federal da Para\'\i ba, 58051-970 Jo\~ao Pessoa, PB, Brazil}

\begin{abstract}
We study vortices in generalized Maxwell-Higgs models, with the inclusion of a quadratic kinetic term with the covariant derivative of the scalar field in the Lagrangian density. We discuss the stressless condition and show that the presence of analytical solutions help us to define the model compatible with the existence of first order equations. A method to decouple the first order equations and to construct the model is then introduced and, as a bonus, we get the energy depending exclusively on a function of the fields calculated from the boundary conditions. We investigate some specific possibilities and find, in particular, a compact vortex configuration in which the energy density is all concentrated in a unit circle. 
\end{abstract}
\date{\today}
\maketitle

\section{Introduction}
Vortices are localized structures that appear in two spatial dimensions. They are present in many areas of nonlinear science, and were firstly investigated in the context of fluid mechanics \cite{helmholtz,fluidmec}. These objects also appear in type II superconductors \cite{abrikosov} when one deals with the Ginzburg-Landau theory of superconductivity \cite{glvortex} and may also be present as magnetic domains in magnetic materials and in many other applications in condensed matter \cite{mag,fradkin}.

In high energy physics, in particular, vortices firstly appeared in the Nielsen-Olesen work \cite{NO}, which is perhaps the simplest relativistic model that supports these structures. The model consists of a Maxwell gauge field minimally coupled to a complex scalar field under the Abelian $U(1)$ symmetry in the $(2,1)$ Minkowski spacetime. An interesting feature of the Nielsen-Olesen vortices is that they are electrically neutral and engender quantized magnetic flux. Their equations of motion are of second order and present couplings between the fields. To simplify the problem, first order equations that solve the equations of motion were found in Refs.~\cite{B,VS}. In this case, the first and second order equations are only compatible if the potential is of the Higgs type, a $|\vphi|^4$ potential that engenders spontaneous symmetry breaking. It is worth mentioning that, even with the Bogomol'nyi procedure, the analytical solutions that describe the vortices remain unknown.

Vortices have also been investigated in generalized models with distinct motivations in several works; see, e.g., Refs.~\cite{g0,g1,babichev2,babi3,g2,g3,g4,g5,g6,g7,g8,g9,g10,g11,vtwin,compvortex,godvortex,anavortex}. In particular, k-vortices, which are vortices in models with generalized kinematics, similar to the models studied before in Refs.~\cite{kinf,cosm1,cosm2,babichev1}, were investigated in \cite{babichev2,babi3}, without the presence of a first order formalism and analytical solutions, but with the search for new effects. Another motivation relies on the possibility of specifying the form of potential, imposed by the first order formalism. For instance, in Ref.~\cite{compvortex}, modifications in the magnetic permeability allowed to develop a route to make the vortex compact. Also, in Ref.~\cite{anavortex}, we have developed a method to obtain vortices and to construct a class of models that supports analytical solutions. Recently, in Ref.~\cite{intvortex}, we have found vortices with internal structure, which arise in generalized models with the magnetic permeability controlled by the addition of a neutral field, enlarging the $U(1)$ symmetry to become $U(1)\times Z_2$.

Motivated by the several works that appeared with generalized dynamics, we have developed a first order formalism for these models in Ref.~\cite{godvortex}. This investigation focused on the search for the conditions that could lead to first order equations in a case similar to the one considered before in Ref.~\cite{babichev2}, with the inclusion of a quadratic kinetic term that involves the covariant derivative of the scalar field in the Lagrangian density. In the current work we further explore the subject, extending the previous results of Refs.~\cite{anavortex,godvortex} to this much harder class of models. The main results show how the presence of analytical solutions can be used to construct the model, if one imposes that its equations of motion are solved by solutions of first order differential equations compatible with the stressless condition.

Although we are working in the $(2,1)$ dimensional spacetime with the Minkowski metric, we think that the results of the current work are also of interest to General Relativity (RG), in particular to the case of the so-called Ricci-based theories of gravity (RBG) formulated within the metric-affine approach. For instance, in the recent work \cite{RBG}, the authors unveiled an interesting correspondence between the space of solutions of RBG and RG, under certain circumstances. The results show that it is sometime possible to map complicated nonlinear models into simpler ones, and we think that the models to be explored in the current work can provide novel possibilities of current interest to the scenario explored in \cite{RBG,RBG1}.

To study the subject, the work is organized in a way such that in Sec.~\ref{secmodel} we present the model and the procedure, showing the requirements to make it work in the presence of first order equations. In Sec.~\ref{secexamples}, we illustrate our findings with some new models that support analytical solutions. In particular, we also calculate the magnetic field, energy density and total energy of the vortex analytically, and investigate the possibility of building compact solutions. Finally, in Sec.~\ref{secconclusions} we end the work with some conclusions and an outlook for future investigations.

\section{Model and Procedure}\label{secmodel}

We consider the generalized action $S=\int d^3x \LL$ for a complex scalar field $\vphi$ coupled to a gauge field $A_\mu$ under the local $U(1)$ symmetry in a three-dimensional Minkowski spacetime with metric tensor $\eta_{\mu\nu} = \text{diag}(+,-,-)$. The Lagrangian density to be investigated has the form
\be\label{lx2}
\LL = K(|\vphi|) X - Q(|\vphi|) X^2 + P(|\vphi|)Y - V(|\vphi|).
\ee
In the above expression, $K(|\vphi|)$, $Q(|\vphi|)$ and $P(|\vphi|)$ are non negative functions that modify the dynamics of the model and $V(|\vphi|)$ is the potential. The minus sign in the $X^2$ term is to keep the vortex energy non negative. Also, $X$ and $Y$ defines the kinetic terms of the scalar and gauge fields, respectively, as
\be\label{XY}
X= \overline{D_\mu \vphi}D^\mu \varphi  \quad \text{and}\quad Y=-\frac14 F_{\mu\nu}F^{\mu\nu},
\ee
where $D_\mu = \partial_\mu +ieA_\mu$, $F_{\mu\nu}=\partial_\mu A_\nu - \partial_\nu A_\mu$ and the overline stands for the complex conjugation. The equations of motion for this model are
\bes\label{eomx2}
\begin{align}
& D_\mu (KD^\mu\vphi) - 2D_\mu (QX D^\mu\vphi) + \nonumber \\
&+ \frac{\vphi}{2|\vphi|}\left(-K_{|\vphi|}X+Q_{|\vphi|}X^2 -P_{|\vphi|}Y + V_{|\vphi|}\right) = 0,
 \\ \label{meqsx2}
&\partial_\mu \left(P F^{\mu\nu} \right) = J^\nu,
\end{align}
\ees
where $J_\mu$ is the conserved current, given by the expression $J_\mu = ie\left(K - 2QX\right)(\overline{\vphi}D_\mu \vphi-\vphi\overline{D_\mu\vphi})$. Also, we are using the notation $V_{|\varphi|}=\partial{V}/\partial |\varphi|$, etc.

The energy-momentum tensor $T_{\mu\nu}$ for the generalized model \eqref{lx2} is 
\be
\begin{aligned}
T_{\mu\nu} &= P F_{\mu\lambda}\tensor{F}{^\lambda_\nu}+ (K - 2QX)\left( \overline{D_\mu \vphi}D_\nu \vphi + \overline{D_\nu \vphi}D_\mu \vphi\right) \nonumber\\
           &\hspace{4mm} - \eta_{\mu\nu} \LL.	
\end{aligned}
\ee
We then consider static configurations, take $A_0=0$ and work with the usual ansatz for vortices
\bes\label{ansatz}
\begin{align}
\vphi(r,\theta)&=g(r)e^{i n\theta}, \\
A_i&=\epsilon_{ij} \frac{x^j}{er^2}\left(n-a(r)\right),
\end{align}
\ees
in which $r$ and $\theta$ are the polar coordinates and $n=\pm1,\pm2,\ldots$ is the vorticity. The boundary conditions for $g(r)$ and $a(r)$ are
\bal\label{bcond}
g(0) &= 0, & a(0)&= n,\\
\lim_{r\to\infty} g(r) &= v, & \lim_{r\to\infty} a(r) &= 0,
\eal
where $v$ is the symmetry breaking parameter which is supposed to be present in the model under investigation. The ansatz \eqref{ansatz} makes $X$ and $Y$ to be written as
\be\label{XYm}
X=-{g^\prime}^2 - \frac{a^2g^2}{r^2} \quad\text{and}\quad Y=-\frac{{a^\prime}^2}{2e^2r^2},
\ee
where the prime denotes the derivative with respect to $r$. The magnetic field is given by $B=-F^{12}=-a^\prime/(er)$. This can be used to show that the magnetic flux $\Phi=2\pi\int_0^\infty r dr B(r)$ is quantized, that is,
\be\label{mflux}
\Phi=\frac{2\pi n}{e}.
\ee
The ansatz \eqref{ansatz} can be plugged in the equations of motion \eqref{eomx2}, which take the form
\bes\label{eomansatzx2}
\begin{align}\label{secansatzg}
\frac{1}{r} \left(r(K - 2QX) g^\prime\right)^\prime -\frac{(K - 2QX) a^2g}{r^2} &\nonumber\\
- \frac12 (-K_gX + Q_g X^2 -P_gY + V_g) &= 0, \\ \label{eomaansatzx2}
r\left(P\frac{a^\prime}{er} \right)^\prime - 2e(K - 2QX) ag^2 &= 0,
\end{align}
\ees
where $K_g=\partial K/\partial g,$ etc. The components of the energy-momentum tensor are
\bes
\begin{align}
T_{00} &= -KX + Q X^2 - PY + V, \label{rho}\\
T_{12} &= (K - 2QX) \left( {g^\prime}^2 - \frac{a^2g^2}{r^2} \right) \sin(2\theta), \\ 
T_{11} &= P \frac{{a^\prime}^2}{e^2r^2} + 2(K - 2QX) \left({g^\prime}^2\cos^2\theta+\frac{a^2g^2}{r^2}\sin^2\theta \right)\nonumber \\&+ \LL, \\ 
T_{22} &= P \frac{{a^\prime}^2}{e^2r^2} + 2(K - 2QX) \left({g^\prime}^2\sin^2\theta+\frac{a^2g^2}{r^2}\cos^2\theta \right)\nonumber \\& + \LL.
\end{align}
\ees
Up to this point, the scenario is quite similar to the one investigated before in Ref.~\cite{babichev2}. Here, however, we want to go further and search for a first order framework that help us to find analytical solutions. We then follow Ref.~\cite{godvortex} and take the stressless conditions, $T_{ij}=0$, which ensure stability of the solution under radial rescaling. This requires the solutions to obey the following first order equations
\bes\label{foeq}
\bal
g^\prime &= \pm \frac{ag}{r},\\
-\frac{a^\prime}{er} &= \pm \sqrt{\frac{2(V-QX^2)}{P}}\label{foeq2}.
\eal
\ees
They allow us to write $X=-2{g^\prime}^2 = -2a^2g^2/r^2$. The above equations, however, must be compatible with the equations of motion \eqref{eomansatzx2}. Similarly to the case that was shown in Ref.~\cite{godvortex}, for $K(|\vphi|)=0$ and $Q(|\vphi|)$ constant, this requirement leads to a constraint that depends on $a$, $g$ and $r$. Therefore, it is hard to obtain a constraint in terms of $g$ and reconstruct the model by finding the explicit form of the potential in terms of $K(|\vphi|)$, $Q(|\vphi|)$ and $P(|\vphi|)$, as in the case $Q(|\vphi|)=0$ that was carefully investigated in Ref.~\cite{anavortex}. The main issue appears because $X$ does not depend exclusively on $g$, but also on $a$ and $r$; see Eq.~\eqref{XYm}. Nevertheless, if the analytical solutions, as well as their inverses, are known, we may write $X$ exclusively in terms of $g$, which we call $X(g)$. By substituting Eqs.~\eqref{foeq} into Eq.~\eqref{eomaansatzx2}, the following constraint arises
\be\label{constraint}
\frac{d}{dg}\sqrt{2P(V-QX^2(g))}=-2eg(K - 2QX(g)).
\ee
One may wonder if the compatibility of Eqs.~\eqref{foeq} with Eq.~\eqref{secansatzg} does not imply into another constraint. Nonetheless, as it was demonstrated in Ref.~\cite{godvortex}, once the above constraint is satisfied and the solutions solve Eqs.~\eqref{foeq}, Eq.~\eqref{secansatzg} becomes an identity. In our model, the choice of the functions $P(g)$, $Q(g)$ and $K(g)$ must be done in a way that it allows the symmetry breaking of the potential $V(g)$ to match with the boundary conditions in Eq.~\eqref{bcond}.

The energy density $\rho=T_{00}$ is given by Eq.~\eqref{rho}. By using the first-order equations \eqref{foeq}, it can be written as
\be\label{rhosimp}
\begin{split}
\rho &=P(g)\frac{a^{\prime\, 2}}{e^2r^2} + 2K(g){g^\prime}^2+8Q(g){g^\prime}^4\\
     &= 2V(g)-K(g) X(g).
\end{split}
\ee
Here, we follow the procedure developed in Ref.~\cite{godvortex} and introduce an additional function $W(a,g)$, defined by
\bal
W_a &=P\frac{a^\prime}{e^2r},\\
W_g &=2(K-2QX(g))rg^\prime,
\eal
where $W_g=\partial W/\partial g$ and $W_a=\partial W/\partial a$. By combining the first order equations \eqref{foeq} and the constraint \eqref{constraint}, one can show that
\be\label{Wag}
W(a,g)=-\frac{a}{e}\sqrt{2P(V-QX^2(g))}.
\ee
In this case, we can write the energy density as
\be
\rho=\frac1r\frac{dW}{dr},
\ee
which can be integrated all over the plane to provide the energy
\be
\begin{split}
	E &= 2\pi|W(a(\infty),g(\infty))-W(a(0),g(0))|,\\
      &= 2\pi|W(n,0)|.
\end{split}
\ee

Now, we follow the route suggested in Ref.~\cite{anavortex} and develop a procedure to build analytical solutions. This can be achieved by decoupling the first order equations \eqref{foeq}, as we describe below. For simplicity, we consider dimensionless fields and take $e,v=1$; also, we work with unity vorticity, setting $n=1$, which means to consider only the upper signs in Eq.~\eqref{foeq}.

In order to decouple the first order equations, we introduce the generating function $R(g)$ such that
\be\label{R}
r\frac{dg}{dr}=R(g).
\ee
Therefore, for a given $R(g)$, we can solve the above equation and obtain $g(r)$ obeying the boundary conditions
\eqref{bcond}. By using this into Eqs.\eqref{foeq}  we obtain
\be\label{a}
a(r)=\frac{R(g(r))}{g(r)}.
\ee

We also introduce another function, $M(g)$, which is defined by $M(g) = -\sqrt{2\left(V(g)-Q(g)X^2(g)\right)/P(g)}$. By using this and the constraint in Eq.~\eqref{constraint}, we get
\bes\label{VGw}
\bal
V(g)&=\frac12 P(g)M^2(g)+Q(g)X^2(g),\\
K(g)&=\frac1{2g}\frac{d}{dg}\left(P(g)M(g)\right)+2\,Q(g)X(g).
\eal
\ees
One can show that $M(g)$ is obtained in terms of the given function $R(g)$ from Eq.~\eqref{foeq2}:
\be\label{M}
M(g)=\frac{R(g)}{q^2(g)}\frac{d}{dg}\left(\frac{R(g)}{g}\right),
\ee
where $q(g)$ is the inverse of $g(r)$. This procedure is valid if $X$ is written only as a function of $g$. Using the definition in Eq.~\eqref{R}, we find
\be\label{Xg}
X(g)=-2R(g)^2/q^2(g).
\ee
We can also take advantage of the function $M(g)$ to write the magnetic field as
\be\label{BM}
B(r)=-M\big(g(r)\big),
\ee
and Eq.~\eqref{Wag} as $W(a,g)= a\,P(g)\,M(g)$, which leads to the total energy
\be\label{eny1}
E=-2\pi\,P(0)M(0).
\ee
This procedure decouples the first-order equations in a manner that the solutions depend only on the generating function $R(g)$. As $M(g)$ depends only on $R(g)$ and $q(g)$, we see from Eqs.~\eqref{VGw} that we have two equations that constrain the functions $V(|\vphi|)$, $P(|\vphi|)$, $K(|\vphi|)$ and $Q(|\vphi|)$. This means that there are several models that support the same analytical solutions defined by Eq.~\eqref{R}. Therefore, to find the explicit form of the models, we need to suggest two of the aforementioned functions. Even though these functions lead to the same solutions and magnetic field, they modify the energy density in Eq.~\eqref{rhosimp}. Thus, one must choose functions that lead to a well defined energy.

We also highlight here that the above procedure to construct the model, described by Eqs.~\eqref{VGw}-\eqref{Xg}, is only valid in the interval $|\vphi|\in[0,1]$, which is the one where the solution exists, according to the boundary conditions \eqref{bcond}. Nonetheless, it is important to suggest non negative functions and a potential that supports a minimum at $|\vphi| = 1$, in order to include spontaneous symmetry breaking and avoid instabilities and negative energies.

\section{Specific examples}\label{secexamples}

Let us now illustrate our procedure with some examples. We firstly suggest an $R(g)$ that leads to analytical solutions and then apply the method in Eqs.~\eqref{VGw}-\eqref{Xg} to construct the model.

\subsection{First example}
\begin{figure}[t!]
\includegraphics[width=4.2cm]{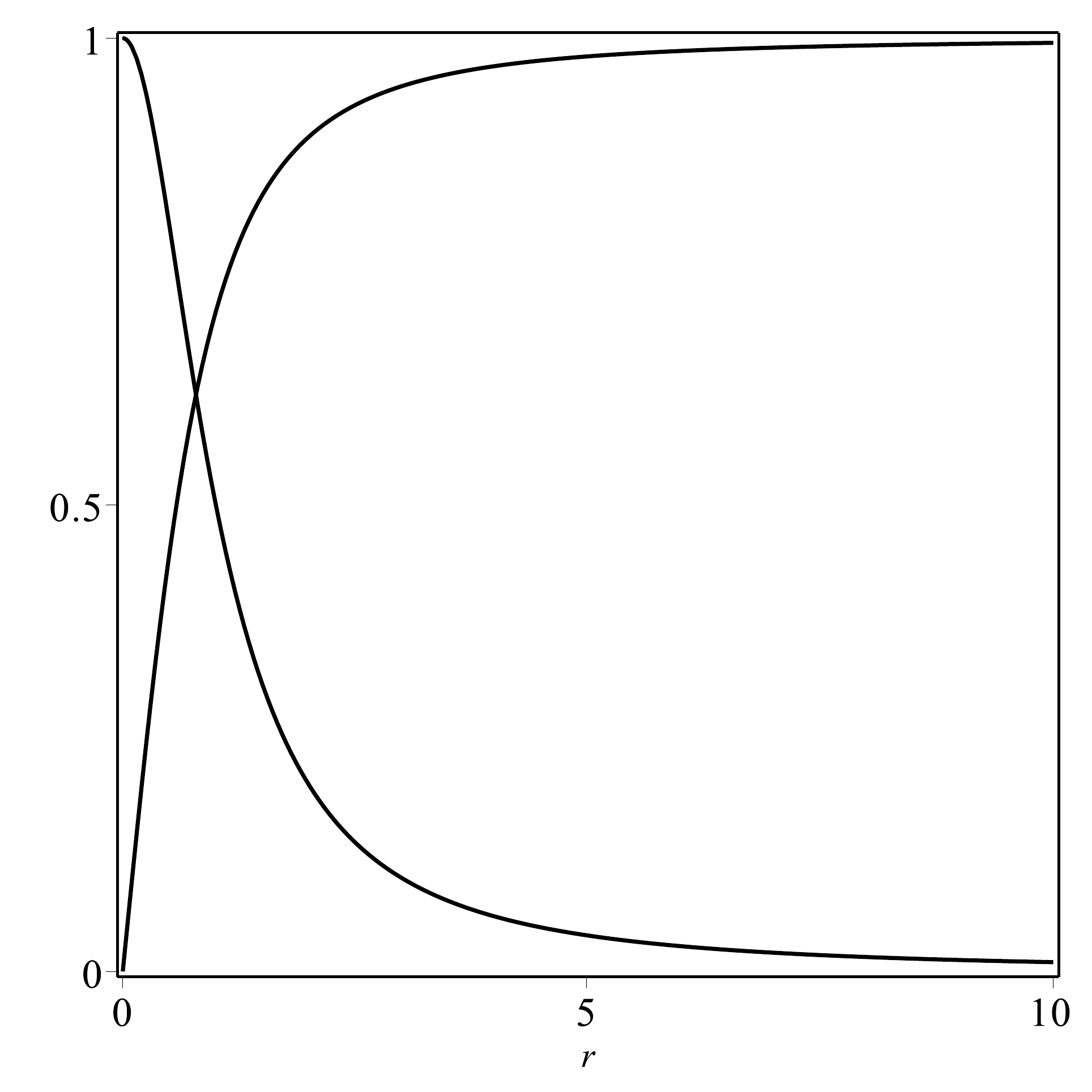}
\includegraphics[width=4.2cm]{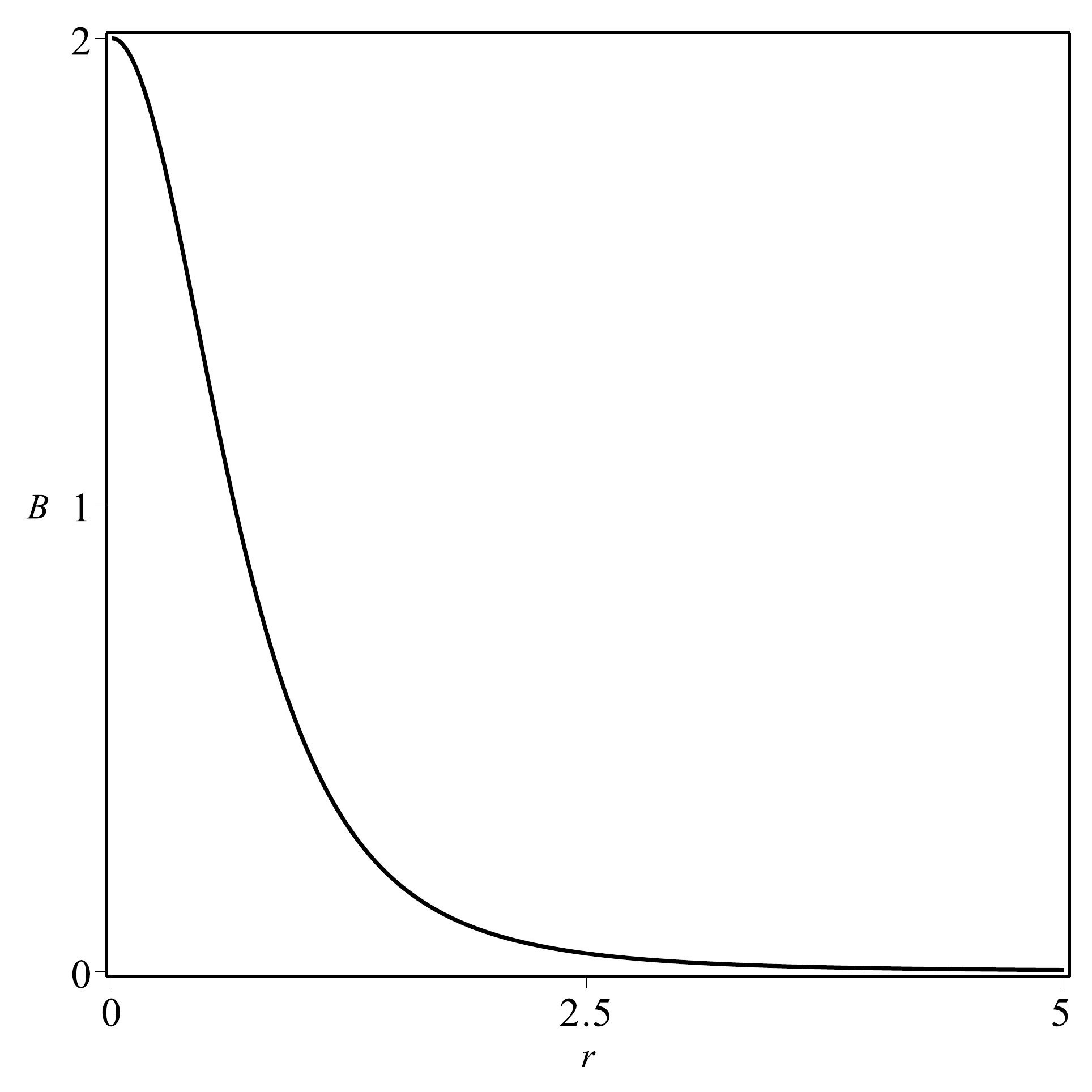}
\caption{In the left panel, we display the solutions $a(r)$ (descending line) and $g(r)$ (ascending line) in Eq.~\eqref{sol1}. In the right panel, we show the magnetic field in Eq.~\eqref{B1}.} 
\label{fig1}
\end{figure}

The first example arises from the generating function
\be\label{R1}
R(g)=g\left(1-g^2\right).
\ee
This function was previously considered in Ref.~\cite{anavortex}, but with a model in which $Q(|\vphi|)=0$, which kills the $X^2$ term in the Lagrangian density. By substituting the above expression in Eqs.~\eqref{R} and \eqref{a} we get the solutions
\be\label{sol1}
g(r)=\frac{r}{\sqrt{1+r^{2}}} \quad\text{and}\quad a(r)=\frac{1}{1+r^{2}},
\ee
which satisfy the boundary conditions \eqref{bcond}. The inverse function of the solution $g(r)$ in Eq.~\eqref{sol1}, combined with Eqs.~\eqref{M} and \eqref{Xg} allow us to write
\bes
\bal
q(g) &= \frac{g}{\sqrt{1-g^2}}\\
M(g) &= -2(1-g^2)^2,\label{M1}\\
X(g) &= -2(1-g^2)^3.\label{X1}
\eal
\ees
Notice that these equations and the solutions in Eq.~\eqref{sol1} are exclusively determined by the function $R(g)$ given in Eq.~\eqref{R1}. This also occurs with the magnetic field, given by Eq.~\eqref{BM}, which leads to
\be\label{B1}
B(r)=\frac2{(1+r^2)^2}.
\ee
In Fig.~\ref{fig1}, we display the solutions \eqref{sol1} and the magnetic field given above. Notice that their behavior is similar to the one for the Nielsen-Olesen case \cite{NO,VS}.

\begin{figure}[!htb]
\includegraphics[width=5cm]{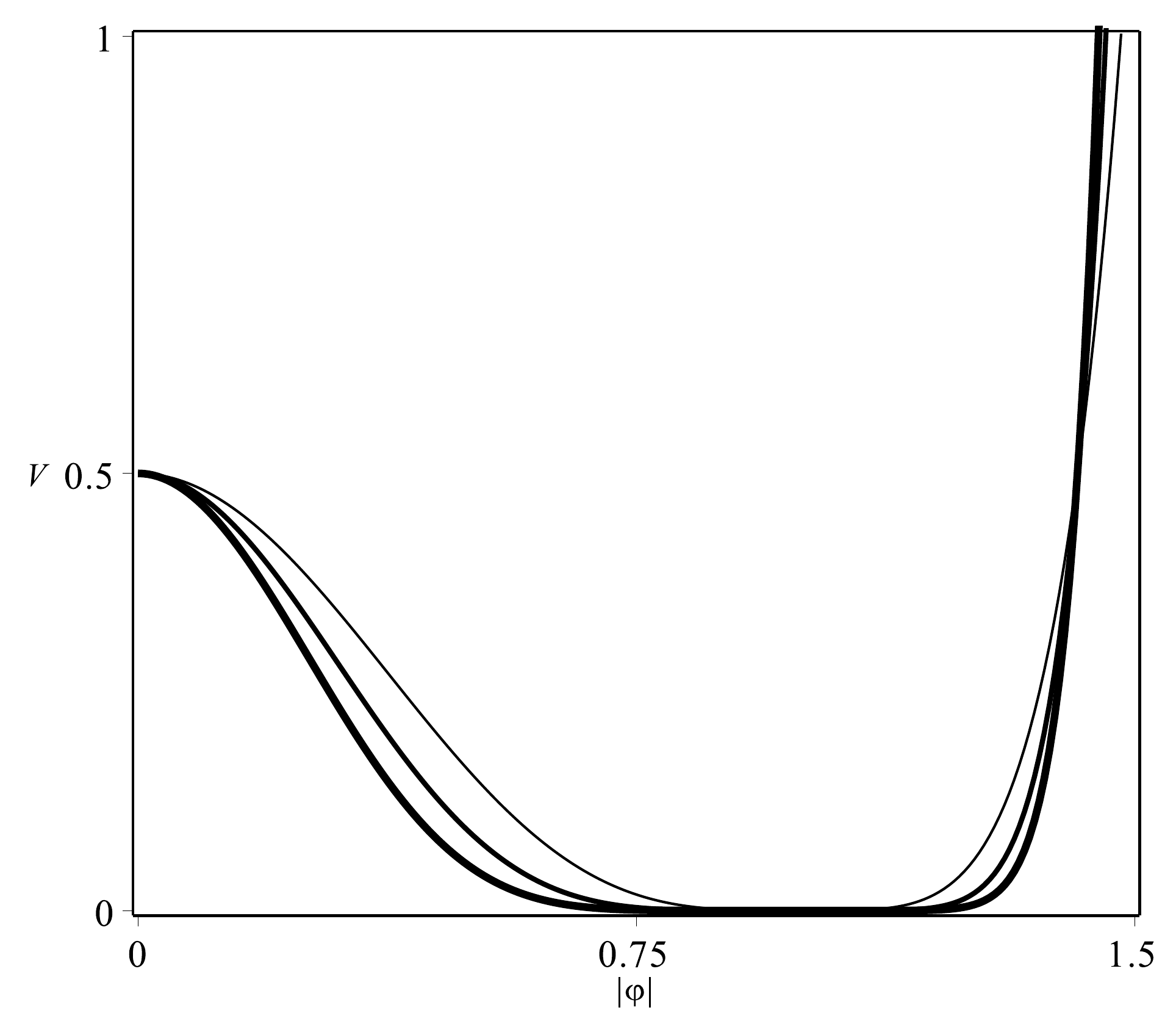} 
\caption{The potential in Eq.~\eqref{V1} for $s=4$, $s=6$ and $s=8$. The thickness of the lines increases with $s$.} 
\label{fig2}
\end{figure}

In order to construct a model that supports the solutions in Eq.~\eqref{sol1}, we use Eqs.~\eqref{VGw}. Firstly, though, we need to suggest an explicit form for two of the functions among $K(|\vphi|)$, $Q(|\vphi|)$, $P(|\vphi|)$ and $V(|\vphi|)$. We consider the potential
\be\label{V1}
V(|\vphi|)=\frac12\left|1-|\vphi|^2\right|^s,
\ee
where $s>2$ is a real number. It presents a set of minima at $|\vphi|=1$ and a local maximum at $|\vphi|=0$ as illustrated in Fig.~\eqref{fig2}.
The other function that we suggest is
\be\label{Q1}
Q(|\vphi|)=\frac{\alpha}{2} \left|1-|\vphi|^2\right|^{s-6},
\ee
where $\alpha$ is a real, non negative parameter. The case investigated in Ref.~\cite{anavortex} is obtained for $\alpha=0$. By substituting the above $Q(|\vphi|)$ and the potential \eqref{V1} in Eqs.~\eqref{VGw} we obtain
\bes\label{PK1}
\bal
P(|\vphi|) &= \frac14\left(1-4\alpha\right)\left|1-|\vphi|^2\right|^{s-4},\\
K(|\vphi|) &= \frac12\left(s-2-4\alpha(s-1)\right)\left|1-|\vphi|^2\right|^{s-3}.
\eal
\ees
In order to avoid negative coefficients in the above functions, we impose the condition $\alpha<(s-2)/4(s-1)$. The functions in Eqs.~\eqref{V1}, \eqref{Q1} and \eqref{PK1} determine the model \eqref{lx2}. We want to emphasize here that this model can only be obtained explicitly because we know the analytical solutions before its construction.

\begin{figure}[!htb]
\includegraphics[width=5cm]{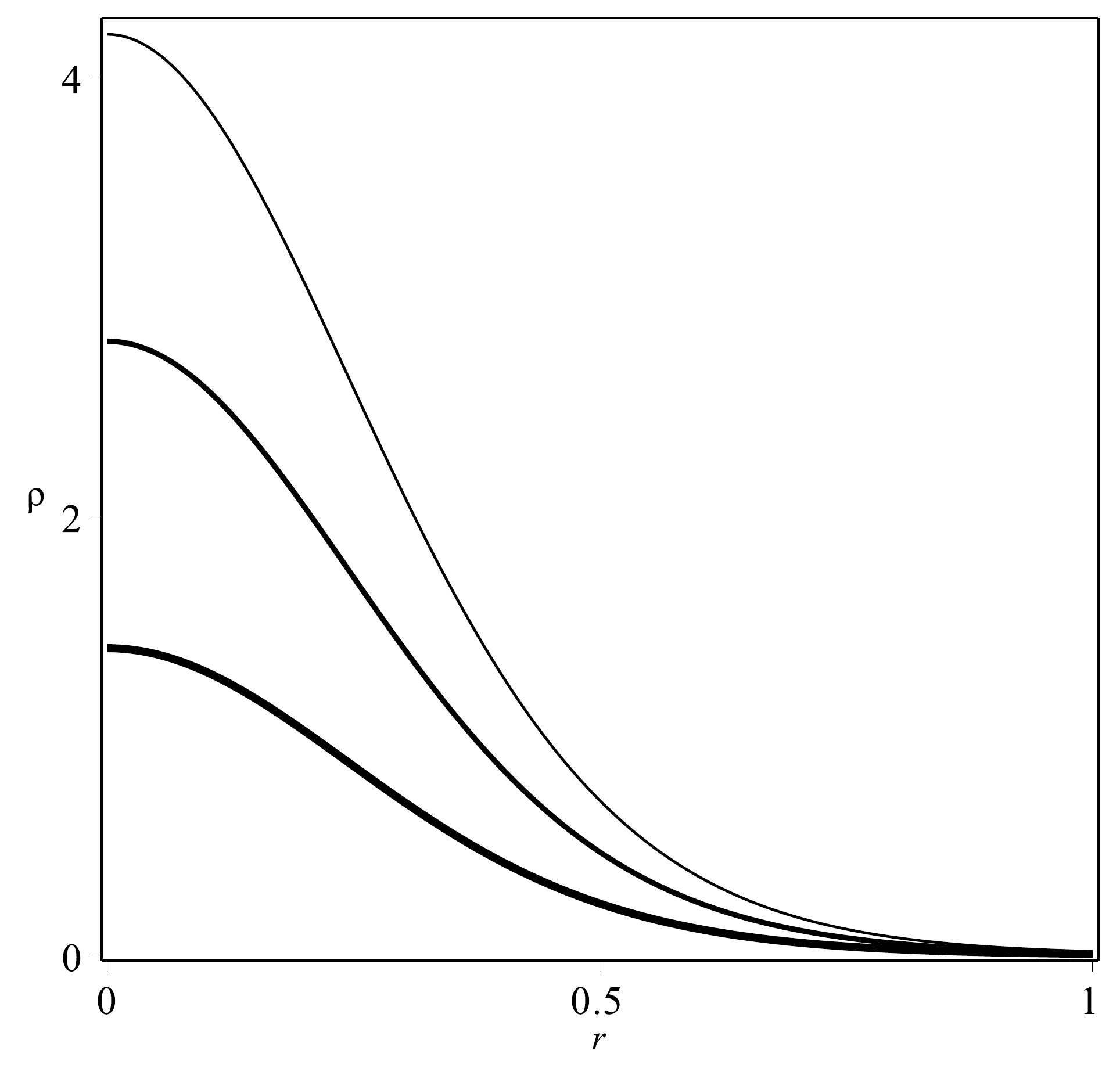} 
\caption{The profile of the energy density in Eq.~\eqref{rho1} for $s=8$ and $\alpha=0.1,0.15$ and $0.2$. The thickness of the lines increases with $\alpha$.} 
\label{fig3}
\end{figure}

The energy density can be calculated from Eq.~\eqref{rhosimp}, which lead us to
\be\label{rho1}
\rho(r)=\frac{(1-4\alpha)(s-1)}{(1+r^2)^s}.
\ee
By a direct integration, one can show that the energy is $E=(1-4\alpha)\,\pi$, which matches with the result obtained by Eq.~\eqref{eny1}. Notice that only the parameter $\alpha$ modifies the energy. The above energy density can be seen in Fig.~\ref{fig3}.

Another models can be generated straightforwardly from the same choice of $R(g)$ in Eq.~\eqref{R1}, that present well defined $V(|\vphi|)$, $P(|\vphi|)$, $Q(|\vphi|)$ and $K(|\vphi|)$ for all $\vphi$.

\subsection{Second example}
\begin{figure}[t!]
\includegraphics[width=4.2cm]{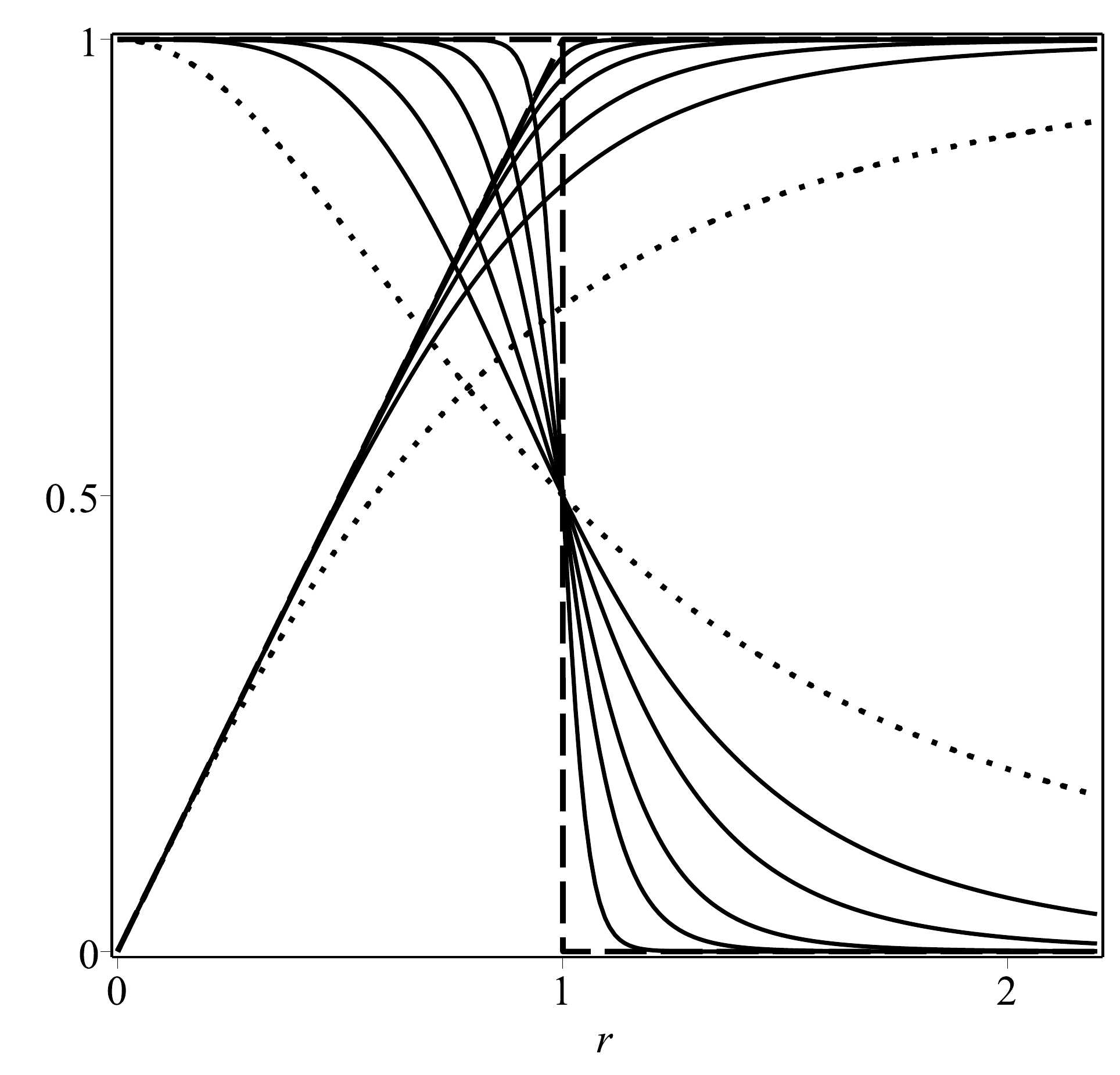}
\includegraphics[width=4.2cm]{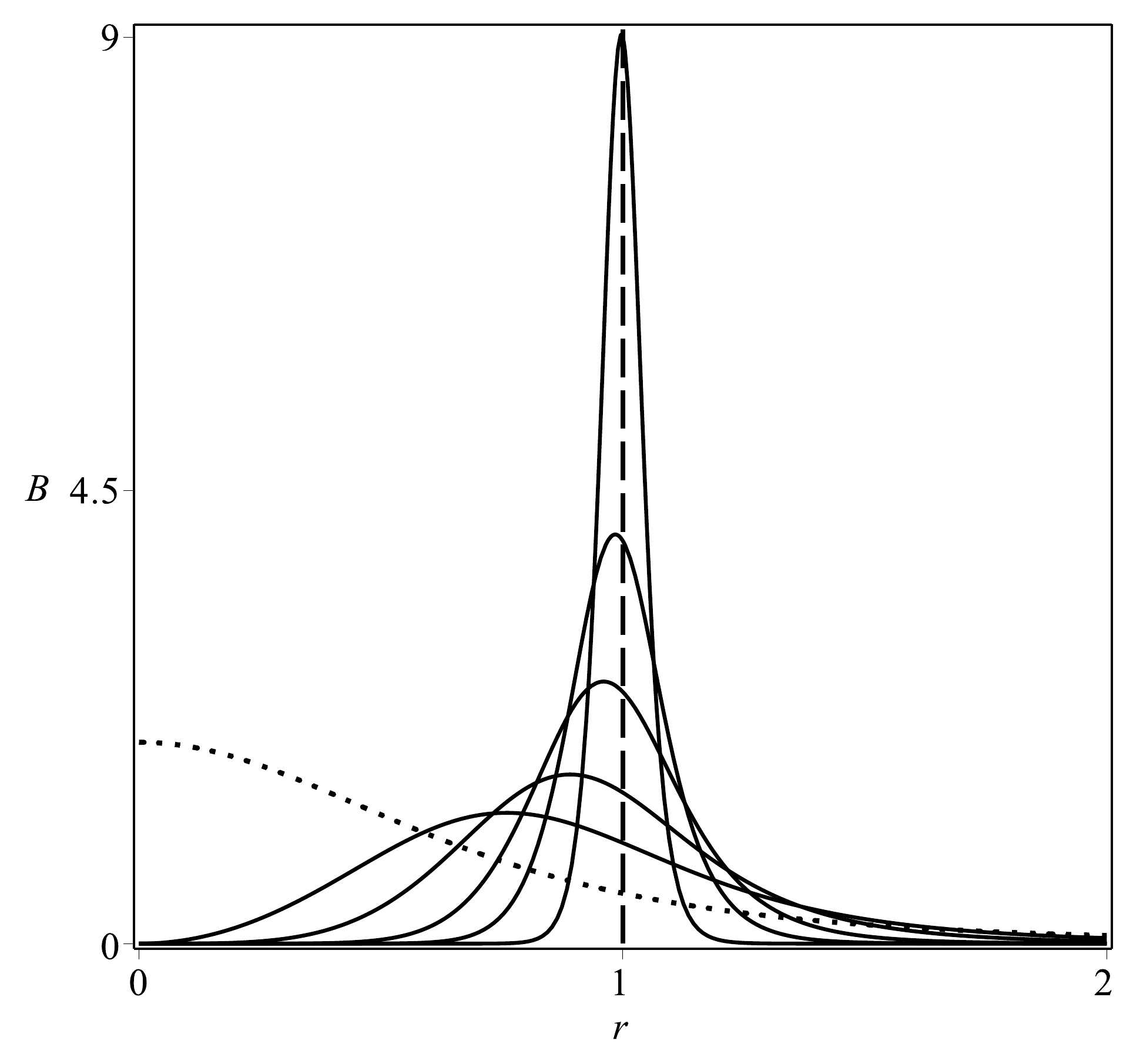}
\caption{In the left panel, we display the solutions $a(r)$ (descending lines) and $g(r)$ (ascending lines) in Eq.~\eqref{sol2}. In the right panel, we show the magnetic field in Eq.~\eqref{B2}. The dotted lines represent the case $l=1$ and the dashed ones stand for the compact limit in Eqs.~\eqref{solc} and \eqref{bc}.} 
\label{fig4}
\end{figure}

Here, we consider a generalization of the previous example by considering the generating function to be
\be\label{R2}
R(g)=g\left(1-g^{2l}\right),
\ee
where $l$ is a non negative real parameter. This function was also investigated in Ref.~\cite{anavortex}, but with $Q(|\vphi|)=0$. From Eqs.~\eqref{R} and \eqref{a}, we get the analytical solutions
\be\label{sol2}
g(r)=\frac{r}{(1+r^{2l})^{1/2l}}, \quad\text{and}\quad a(r)=\frac{1}{1+r^{2l}},
\ee
which satisfy the boundary conditions \eqref{bcond}. From the inverse of the solution $g(r)$, combined with Eqs.~\eqref{M} and \eqref{Xg}, we obtain
\bes
\bal
q(g) &= \frac{g}{(1-g^{2l})^{1/2l}}\\
M(g) &= -2g^{2l-2}(1-g^{2l})^{1+1/l},\label{M2}\\
X(g) &= -2(1-g^{2l})^{2+1/l}.\label{X2}
\eal
\ees
As in the previous model, these equations and the solutions in Eq.~\eqref{sol2} are solely determined by the $R(g)$ in Eq.~\eqref{R1}. The same is valid for the magnetic field in Eq.~\eqref{BM}, which leads to
\be\label{B2}
B(r)=\frac{2lr^{2l-2}}{(1+r^{2l})^2}.
\ee
One can show that, as $l$ increases, the solutions in Eqs.~\eqref{sol2} tend to compactify
\bes\label{solc}
\bal
a_c(r) &=
\begin{cases}
1,\,\,&r\leq 1\\
0, \,\, & r>1,
\end{cases}  \\
g_c(r) &=
\begin{cases}
r,\,\,\,&r\leq 1\\
1, \,\,\, & r>1,
\end{cases}
\eal
\ees
The same happens for the magnetic field in Eq.~\eqref{B2}, which for very large $l$ tends to
\be\label{bc}
B_c(r) = \frac{\delta(r-1)}{r},
\ee
where $\delta(z)$ is the Dirac delta function. In Fig.~\ref{fig4}, we depict the solutions \eqref{sol2} and the magnetic field given above for several values of $l$, including the compact limit in Eq.~\eqref{solc}.

\begin{figure}[!htb]
\includegraphics[width=5cm]{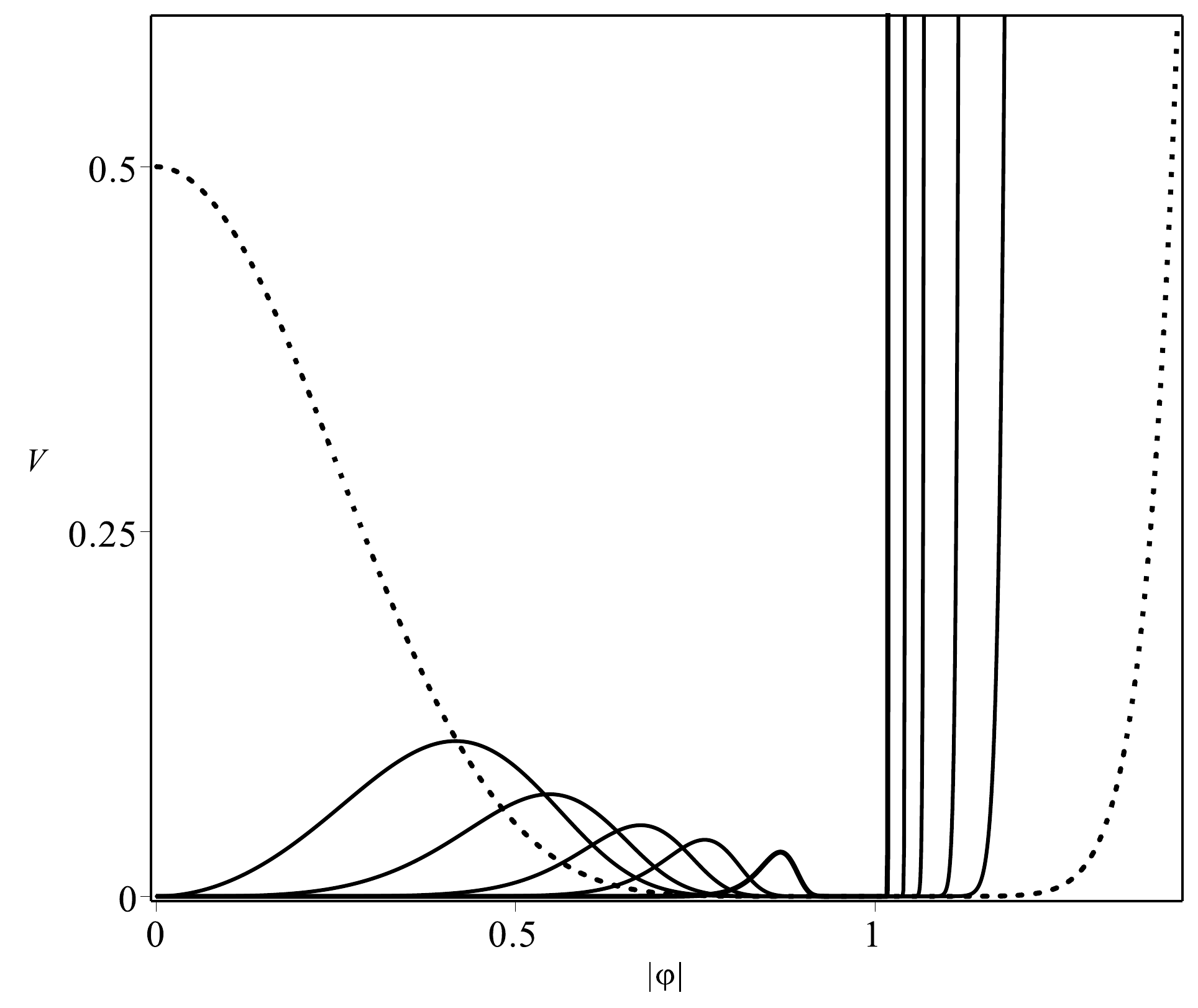} 
\caption{The potential in Eq.~\eqref{V2} for $\beta=8$ and several values of $l$. The dotted line stands for the case $l=1$.} 
\label{fig5}
\end{figure}

Again, to find the functions $K(|\vphi|)$, $Q(|\vphi|)$, $P(|\vphi|)$ and $V(|\vphi|)$ we must suggest two of them and use Eqs.~\eqref{VGw}. We take the potential in the form
\be\label{V2}
V(|\vphi|)=\frac12 l |\vphi|^{2l-2}\left|1-|\vphi|^{2l}\right|^{\beta l},
\ee
where $\beta>2$ is a real number. This potential presents minima at $|\vphi|=1$ for any $l$. The point $|\vphi|=0$ is a maximum for $l=1$ and a minimum for $l>1$. This behavior is shown in Fig.~\eqref{fig5}.
Together with the potential in Eq.\eqref{V2}, we keep the same lines of the previous example and suggest the $X^2$ term in the Lagrangian density to be modified by 
\be
Q(|\vphi|)=\frac12 \alpha l\,  |\vphi|^{2l-2}\left|1-|\vphi|^{2l}\right|^{\beta l-4-2/l},
\ee
where $\alpha>0$ is a real parameter. Substituting $V(|\vphi|)$ and $Q(|\vphi|)$ in Eqs.~\eqref{VGw}, we obtain
\bal
P(|\vphi|)&=\frac1{4l}(1-4\alpha)\,|\vphi|^{2-2l}\left|1-|\vphi|^{2l}\right|^{\beta l-2-2/l},\\
K(|\vphi|)&=\frac1{2}\left((1-4\alpha)(\beta l^2-1)-l\right) \nonumber\\
          &\hspace{4mm}\times|\vphi|^{2l-2}\left|1-|\vphi|^{2l}\right|^{\beta l-2-1/l}.
\eal
To avoid the presence of negative coefficients in the above expressions, we impose that $\alpha<(\beta l^2-l-1)/4(\beta l^2-1)$.

\begin{figure}[!htb]
\includegraphics[width=4.2cm]{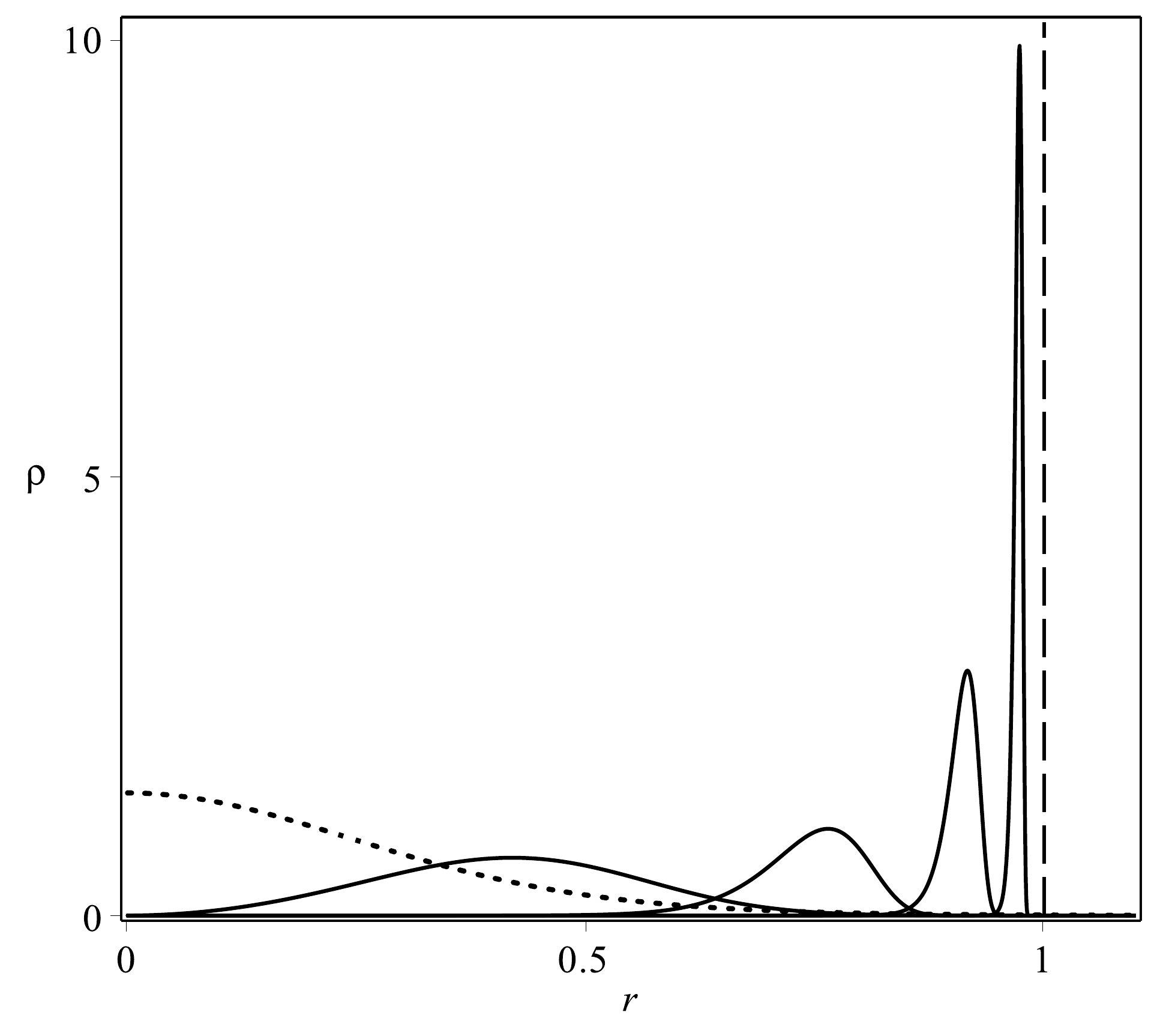}
\includegraphics[width=4.2cm]{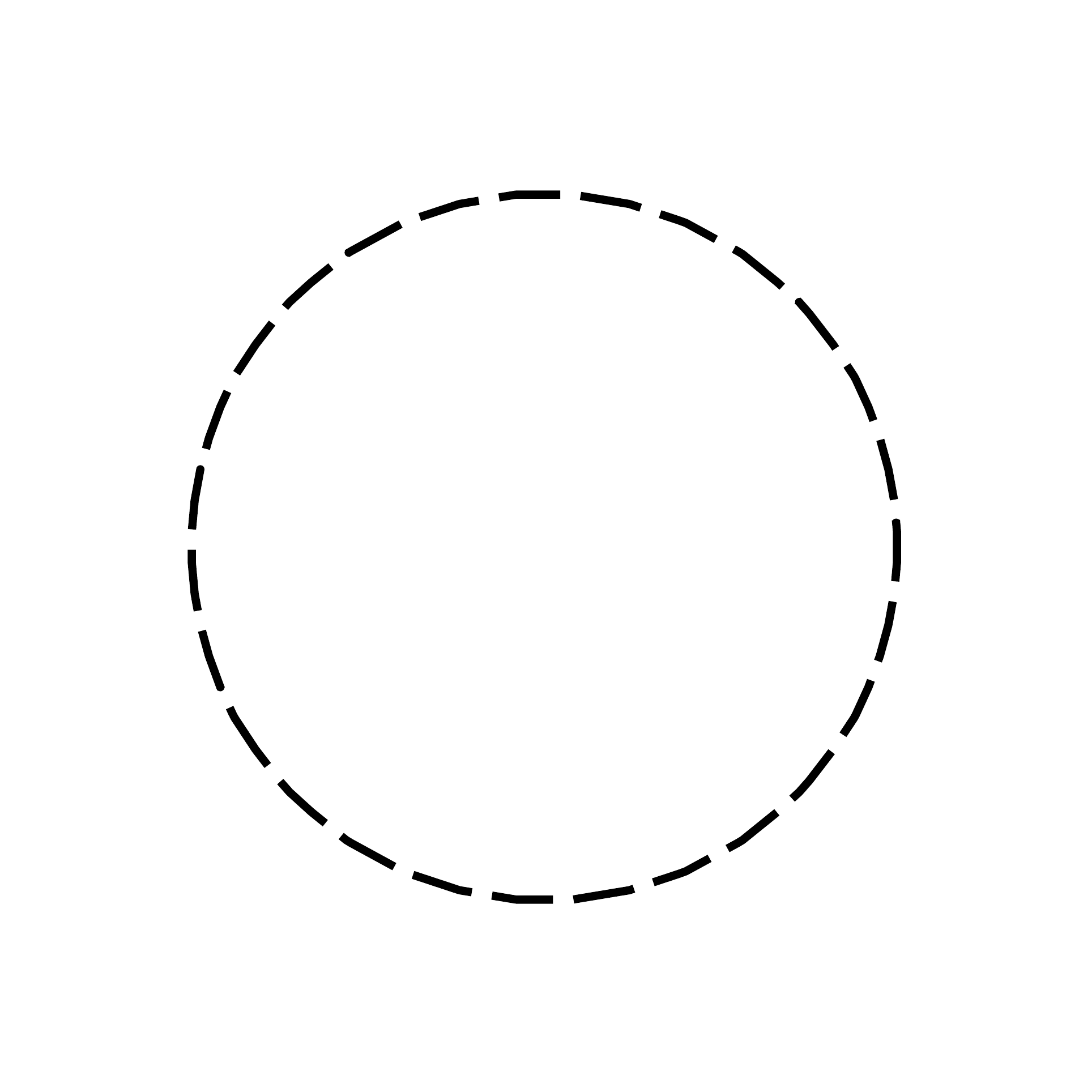} 
\caption{The energy density of Eq.~\eqref{rho2} for $\beta=8$, $\alpha=0.2$ and several values of $l$ (left) and its compact limit, $l\to\infty$, in the plane (right). The dotted line represents the case $l=1$ and the dashed ones stand for the compact limit in Eq.~\eqref{rho2c}.} 
\label{fig6}
\end{figure}

The energy density is calculated from Eq.~\eqref{rhosimp}, which leads to
\be\label{rho2}
\rho(r)=\frac{(1-4\alpha)(\beta l^2-1)r^{2l-2}}{(1+r^{2l})^{\beta l+1-1/l}}.
\ee
One may integrate it to get the total energy $E=(1-4\alpha)\,\pi$, which matches with the value obtained by Eq.~\eqref{eny1}. Again, only the parameter $\alpha$ modifies the energy of the vortices, meaning that the $X^2$ term in the Lagrangian density \eqref{lx2} play a significant role in the model. Following a similar procedure that was done in Ref.~\cite{anavortex}, one can show that the energy density tends to compactify into a ringlike region of unit radius in the plane, described by
\be\label{rho2c}
\rho(r) = \frac12(1-4\alpha)\delta(r-1).
\ee
In Fig.~\ref{fig6}, we display the energy density for several values of $\alpha$, including the compact limit given above. Its behavior, even with the presence of the parameter $\alpha$, is qualitatively similar to the one found in Ref.~\cite{compcs} for the compactification of vortices in a generalized Chern-Simons-Higgs model. 

\section{Comments and Conclusions}\label{secconclusions}

In this work, we have developed a procedure that allow to construct k-vortex models that support a first order framework. As we discussed above, the method is important because the constraint that dictates the form of the potential cannot be solved in general in the presence of the squared kinetic term of the scalar field, $X^2$, in the Lagrangian density. Thus, it seems to be very hard to start from a model with this term and find the potential that leads to the first order equations compatible with the stressless condition, vital to the stability of the system.

Nevertheless, we got inspiration from the recent works \cite{godvortex,anavortex} and noticed that, if an analytical solution is known, we can construct a model that satisfy the stressless condition and find the energy depending exclusively on a function of the fields calculated from the boundary conditions. In order to achieve this, we have introduced the generating function $R(g)$ that decouples the first order equations. It is interesting feature of this procedure that it shows there is a class of models that leads to the same analytical, stressless solutions and their respective magnetic fields, which only depend on the generating function. However, the energy density as well as the total energy depend on the model to be chosen, so we have to properly define the model, to make it bahaves adequately.

It is worth commenting the fact that a similar method can be developed for the more general Lagrangian density $\LL= f(X,|\vphi|) + P(|\vphi|)Y - V(|\vphi|)$. Thus, among the myriad of possibilities, one may develop a construction method for the kinetic term of the scalar field being of the Born-Infeld type, for instance. Other perspectives should include the possibility to consider the case in which the dynamics of the gauge field is driven by the Chern-Simons term, which cannot be multiplied by $P(|\vphi|)$ if one wants to keep gauge invariance. Since the magnetic permeability of the model is generalized, one may also investigate the presence of vortices in metamaterials; see, e.g., Refs.~\cite{M1,M2,M3}. Furthermore, as the model supports the $W$ in Eq.~\eqref{Wag}, one may seek for supersymmetric extensions, to investigate how the supersymmetry works in this scenario to lead us with first order differential equations. One may also try to extend these results to other topological structures, such as monopoles and skyrmions. We hope to report on some of the above issues in the near future. 

\acknowledgements{We would like to acknowledge the Brazilian agency CNPq for partial financial support. DB thanks support from grant 306614/2014-6, LL thanks support from grant 303824/2017-4, MAM thanks support from grant 140735/2015-1 and RM thanks support from grant 306826/2015-1.}\\

\begin{center}
{***}
\end{center}

The authors declare that there is no conflict of interest regarding the publication of this paper.

\end{document}